\documentclass[review]{elsarticle}
\usepackage{amsmath}
\usepackage{bm}
\usepackage{lineno,hyperref}
\usepackage{gensymb}
\usepackage{xcolor}
\usepackage{listings}
\usepackage{multirow}
\usepackage{makecell}
\usepackage{comment}
\definecolor{codegreen}{rgb}{0,0.6,0}
\definecolor{codegray}{rgb}{0.5,0.5,0.5}
\definecolor{codepurple}{rgb}{0.58,0,0.82}
\definecolor{backcolour}{rgb}{0.95,0.95,0.92}

\modulolinenumbers[5]

\journal{Journal of \LaTeX\ Templates}

\begin{document}

\begin{frontmatter}

\title{Effects of stenotic aortic valve \\ on the left heart hemodynamics: \\ a fluid--structure--electrophysiology approach}

\author[address1]{Francesco Viola\corref{mycorrespondingauthor}}
\cortext[mycorrespondingauthor]{Corresponding author}
\ead{francesco.viola@gssi.it}

\author[address2]{Valentina Meschini}
\author[address1,address3,address4]{Roberto Verzicco 
}
\address[address1]{Gran Sasso Science Institute, Italy}
\address[address2]{Department of Mathematics, University of Rome Tor Vergata, Rome, Italy}
\address[address3]{Department of Industrial Engineering, University of Rome Tor Vergata, Rome, Italy}
\address[address4]{PoF group, University of Twente, The Netherlands}

\begin{abstract}
The aortic valve is a three-leaflet passive structure that, driven by pressure differences between the left ventricle and the aorta, opens and closes during the heartbeat to ensure the correct stream direction and flow rate. 
In elderly individuals or because of particular pathologies, the valve leaflets can stiffen thus impairing the valve functioning and, in turn, the pumping efficiency of the 
heart. 
\\ \indent 
Using a multi-physics left heart model accounting for the electrophysiology, the active contraction of the myocardium, the hemodynamics and the related fluid--structure--interaction, we have investigated the changes in the flow features
for different severities of the aortic valve stenosis. 
We have found that, in addition to the increase of the transvalvular pressure drop and of the systolic jet velocity, 
a stenotic aortic valve significantly alters the wall shear stresses and their spatial distribution over the aortic arch and  valve leaflets, which may induce a remodelling process of the ventricular myocardium.
The numerical results from the multi--physics model are fully consistent with the clinical experience, thus further opening the way for computational engineering aided medical diagnostic. 
\end{abstract}


\end{frontmatter}

\section{Introduction}
The heart is a vital organ whose functioning results from the complex interaction of electrophysiology, tissue mechanics and hemodynamics. Despite its remarkable reliability, some of its parts can deteriorate in time, especially on the left side that feeds the systemic circulation and supports the highest pressure differences \cite{hall2010guyton}.
Among various diseases, valves stenosis is one of the most common occurring whenever the valve leaflets stiffen owing to a progressive calcification which impairs their normal operation yielding an increased risk of myocardial infarction and mortality  \cite{auricchio2011computational,carabello2009aortic,conti2010dynamic}.
Valvular heart disease, indeed, accounts for 10\% to 20\% of all cardiac surgical procedures in the western Countries and
approximately two thirds of them are for aortic valve replacement owing to 
aortic stenosis (AS) disease \cite{braunwald}. 
\\ \indent
 The degree of stenosis associated with the symptoms varies among patients and in most cases AS can now be diagnosed before symptoms onset through physical examination and echocardiography, which has become the standard approach for the diagnosis and evaluation of valve disease since it allows the accurate definition of valve anatomy, including the cause and the severity of AS. 
Grading the AS severity is important to predict the clinical outcome, to this aim the European Association of Echocardiography (EAE) together with the American Society of Echocardiography (ASE) \cite{baumgartner2009echocardiographic} propose to measure the AS severity  in the clinical practice using a combination of parameters such as the aortic valve area (AVA), the velocity of the systolic jet and the mean transvalvular pressure drop (TPD). 
In particular, \textit{mild} AS is characterized by an aortic jet velocity comprises between 2.0 and 2.9 m/s and a mean TPD less than 20 mm Hg, typically with with an AVA of 1.5 to 2.0 cm$^2$.
These values make worse up to an aortic jet velocity of 3.0--3.9 m/s, a mean TPD equal to 20 to 39 mm Hg 
and an AVA in between 1.0 and 1.5 cm$^2$ are considered as \textit{moderate}  AS.
Finally, \textit{severe} AS entails a more significant obstruction to LV outflow usually resulting in an aortic jet velocity exceeding 4 m/sec or greater, a mean TPD of at least 40 mm Hg along with an AVA smaller than 1.0 cm$^2$, which is about 25\% the corresponding value in the healthy case.
These guidelines are used to quantify the severity of the AS and the temporal evolution of the disease since it has been observed 
that when mild obstruction is present, worsening of the calcification occurs in almost all patients, with the interval from mild to severe obstruction ranging from less than 5 to more than 10 years \cite{baumgartner2009echocardiographic}. 
\\ \indent 
A deeper understanding of the basic biologic mechanisms leading to valve dysfunction and novel diagnostic methods are needed to optimize
medical interventions and prevent or slow down the disease progression: Computational engineering  is becoming an added value in medical research since it provides a non invasive framework to study the alterations caused by heart pathologies and to predict the most favorable medical treatment. Numerical simulations allow, indeed, to test and compare new prostheses and surgical procedures and provide a complete access to hemodynamics data, which is unattainable through in-vivo experiments.
For instance, the rate of calcification of both the normal tricuspid and the pathologic bicuspid aortic valves has been investigated employing structural finite--element simulations of the valve leaflets  \cite{weinberg2008multiscale}, whereas
the initiation and growth of calcifications has been investigated re--creating numerically the different calcification growth stages observed in patient-specific tomographic scans \cite{halevi2015progressive}.
On the other hand, a patient-specific computational model to quantify the biomechanical interaction between the transcatheter aortic valve (TAV) stent and the stenotic aortic valve during TAV intervention has been developed to plan pre-operative strategies and facilitate next generation device design \cite{wang2012patient,sturla2016impact}.
Several works than focused on the complex fluid--structure--interaction (FSI) governing the opening and closing dynamics of the aortic valve leaflets in physiological conditions in the case of biological \cite{hsu2014fluid,hsu2015dynamic} and mechanical \cite{de2009direct,DeTullio2011} prosthetic aortic valves. More recently the FSI of calcific aortic valves has been solved both using an Arbitrary Lagrangian-Eulerian (ALE) method \cite{bahraseman2016fluid}  and an immersed boundary (IB) approach \cite{mittalstenosis2018}, where the aorta is modeled as a curved rigid pipe with a 180  turn and three different stenoses with area reductions were investigated.
\\ \indent 
In all these FSI studies, however,  the pulsatile flow in the aorta is driven by imposing an unsteady inflow at the aortic orifice
and the computational domain only comprises the thoracic aorta, thus overlooking the left ventricle and atrium whose dynamics generates the unsteady inflow itself.
The hemodynamics in the aorta and in the left heart are intrinsically coupled owing to the elliptic nature of the 
governing (Navier-Stokes) equations and only considering the whole left heart hemodynamics allows to solve the transvalvular pressure drop, which is a clinical quantity of paramount importance to quantify AS, as introduced above.
\\ \indent
In this work we present a numerical approach to study the AS and to investigate its effects accounting for the whole left--heart. Specifically, also the electrophysiology driving the electrical depolarization and the subsequent contraction of the muscular tissue is solved and the resulting fluid--structure--electrophysiology interaction (FSEI \cite{fsei}) makes the system self--consistent. As a consequence, the valve leaflets kinematics along with the inflow through the valvular orifices induced by the 
heart chambers expansion/contraction are not imposed but are obtained as a part of the numerical results.
The hemodynamics of the whole system is thus investigated in physiological and pathological conditions to highlight the alterations of the ventricular flow produced by the AS,
which is modelled  as previously proposed for mitral valve stenosis \cite{Meschini2019} 
by progressively increasing the severity of the AS from moderate to acute, according to the current risk stratification \cite{zoghbi2003recommendations}. 
The relevant clinical parameters introduced above to assess AS severity such as the AVA, the systolic jet velocity  and mean TPD are measured numerically and validated against clinical observations before studying the wall shear stresses (WSS), which can not be measured in--vivo directly .
\\ \indent
The paper is structured as follows.
In Section~\ref{sec:problem} the problem configuration and the AS model are presented along with the FSEI computational framework. The results on the alteration on the left--heart hemodynamics induced by AS are shown in Section~\ref{sec:results} and compared against the healthy  valve configuration.
The closing remarks and the perspectives for future studies are given in the final section~\ref{sec:discussion}.

\section{Materials and methods }\label{sec:problem} 
\subsection{Problem configuration and aortic stenosis modelling}
\begin{figure}[t!]
\centering
\makebox[\textwidth][c]{\includegraphics[width=0.99\textwidth]{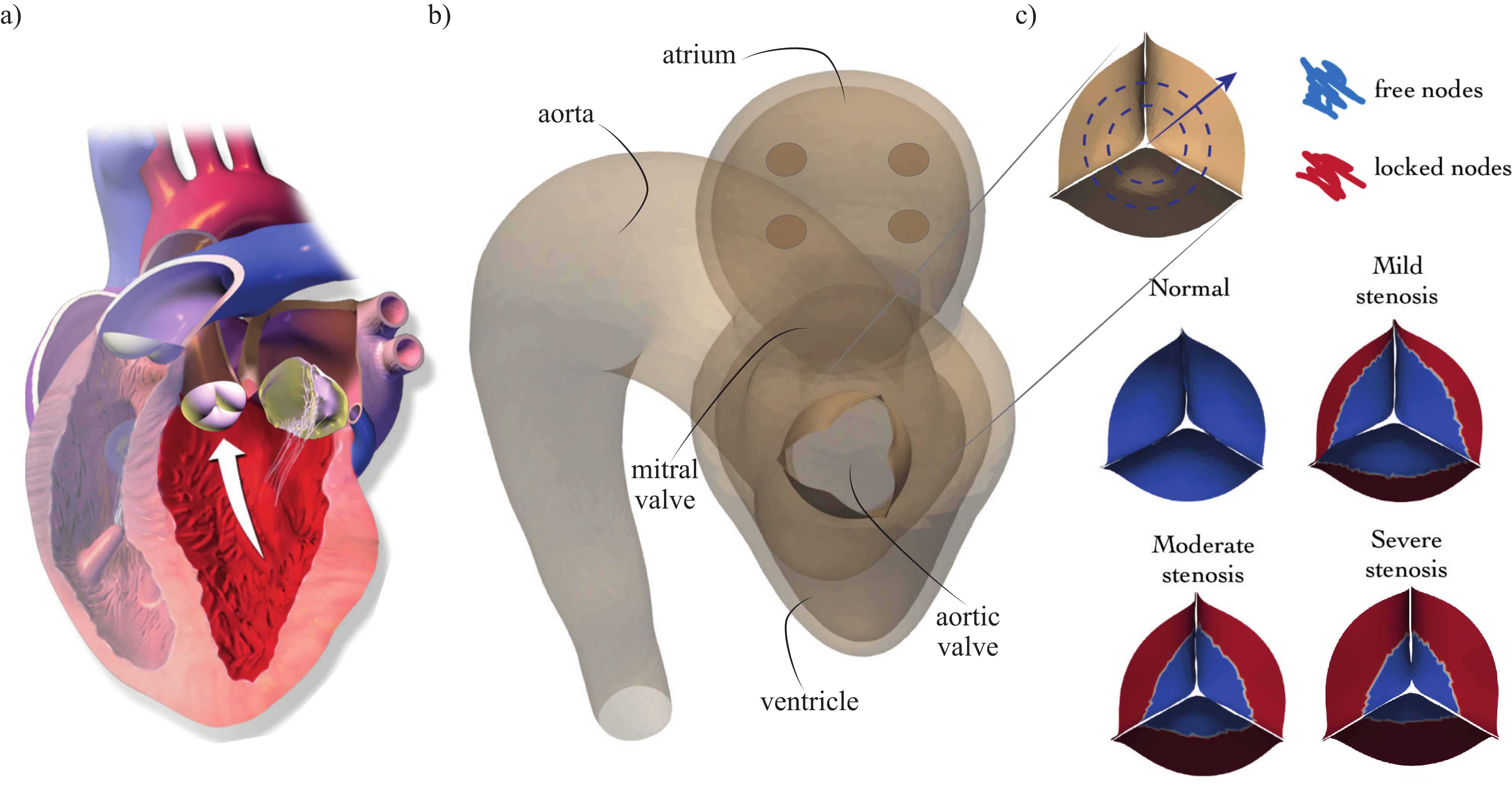}}%
\caption{a) Sketch of the heart highlighting its left pump and corresponding b) left heart geometry used in the computational model comprising ventricle, atrium, thoracic aorta, mitral and aortic valve. c) Reduced mobility regions of the aortic valve leaflets according to the AS model, the blue region indicate healthy tissues whereas the red one calcified tissues.}
\label{fig:EFScoupling}
\end{figure}

We consider the set-up for the left human heart reported in figure~\ref{fig:EFScoupling} 
which reproduces the main geometrical features of the cardiac anatomy as depicted in medical atlas and comprises a left atrium and ventricle having a stress--free volume of 40~ml and 125~ml, respectively. 
The left atrium receives oxygenated blood via four circular inlets corresponding to the inlets of the pulmonary veins and is connected to the left ventricle through the mitral valve, which is made of two leaflets (an anterior and a posterior one) modeled as deformable membranes 2~mm thick mounted on a mitral orifice of diameter $d_m=24$~mm.
During systole the left ventricle contracts and pumps blood through the ascending aorta that bends down 180\degree in the aortic arch and then extends in the thoracic descending aorta. The ventricle and the aorta are connected by the aortic valve that is made of three cusps placed just before the Valsalva sinuses with an aortic orifice of $d_a=20$~mm, with the three leaflets modeled as deformable membranes of thickness 1.5~mm.
The Reynolds number  is defined using as reference length and velocity, the diameter of the atrial orifice and the 
orifice averaged velocity of the aortic jet during peak systole $U_a=1.5$~m/s (see table~\ref{tabella}) corresponding to $Re=U_a d_a/\nu=6250$, with $\nu=4.8 \times 10^{-6}$~m$^2$/s the effective kinematic viscosity for human blood with an hematocrit of $40\%$ (Newtonian blood model). 
\\ \indent 
As mentioned above, the AS depends on the deterioration of the tissues owing to a calcification process yielding a local stiffening over the leaflets, thus reducing their elasticity and mobility. 
In order to reproduce such reduced mobility, AS is modeled by blocking a certain region of the leaflets next to the aortic root 
 (corresponding the red area in figure~\ref{fig:EFScoupling}c) 
 to its initial position as a function of the disease severity, which can be progressively varied from moderate to acute by increasing the calcific area and, concurrently, reducing the AVA. A similar approach was adopted to model correctly the mitral valve stenosis in previous FSI numerical simulations \cite{Meschini2019}. 
Here, we have focused on four different cases: a \textit{normal} valve corresponding to physiological conditions in which the valve leaflets can open  freely according to the hydrodynamic loads; a \textit{mild} stenosis, in which only 30 \% of the orifice area is blocked; a \textit{moderate} stenosis where 55\% of the leaflet area is blocked and a \textit{severe} stenosis corresponding to a calcific obstruction of 80 \% the aortic orifice area.

\subsection{Numerical method: Fluid--Structure--Electrophysiology interaction (FSEI)}\label{sec:fsei}
Building a computational model of the heart entails dealing with the complex fluid--structure interaction between the pulsatile hematic flow and the deforming biological tissues whose active contraction triggered by the electrophysiological system.
Our group has developed a computational framework for solving the full fluid--structure--electrophysiology interaction (FSEI) which has been extensively discussed and validated through a series of studies \cite{Meschini2018,fsei,fseichap} and which is only briefly summarized here.
\begin{figure}[t!]
\centering
\makebox[\textwidth][c]{\includegraphics[width=0.9\textwidth]{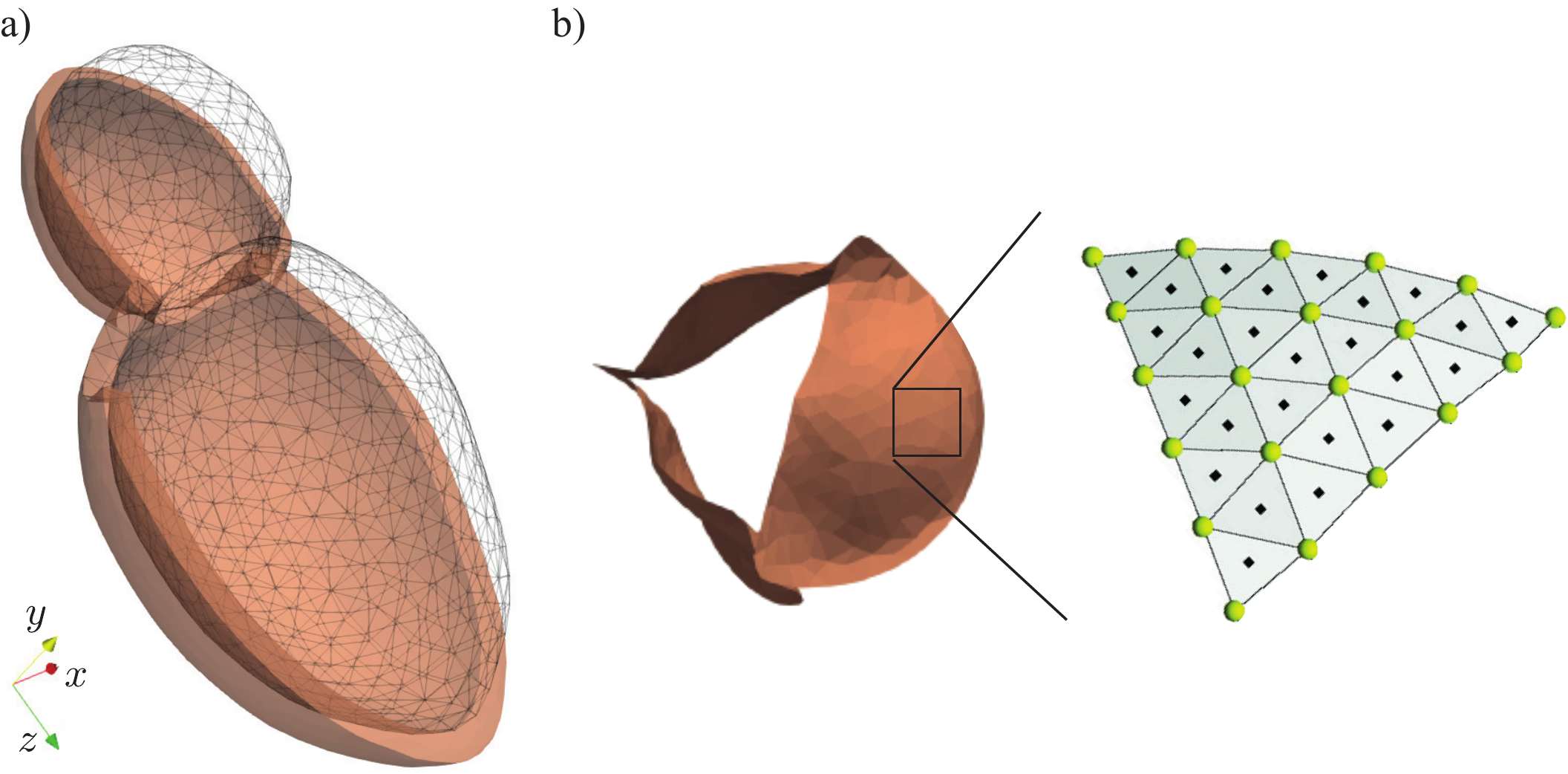}}%
\caption{Snapshots of the a)  3D myocardium and the corresponding 2D endocardium and b) 2D aortic valve. The inset in b) shows a sketch of the Lagrangian markers (black dots) distributed over 2D triangulated meshes.
 }
\label{fig:c2d3d}
\end{figure}
\\ \indent
As shown in Figure~\ref{fig:c2d3d}, a three--dimensional (3D) solver is used for the ventricular and atrial myocardium which are discretized using a tetrahedral mesh, with the endocardium wet by the blood corresponding to a triangular inner surface. On the other hand, thin membranes as the valve leaflets and the aorta are discretized through
two--dimensional (2D) triangulated surfaces. 
The dynamics of the deformable heart tissues is thus solved using a spring--network structural model based on the Fedosov's interaction potential approach  \cite{Fedosov,TuPa} where the mass of the tissues is concentrated at the mesh nodes and an elastic spring is placed 
at each mesh edge connecting two nodes. 
The anisotropic and hyperelastic nature of biological cardiac tissues is modelled by a larger elastic stiffness in the fiber direction, $\hat{\mathbf e}_f$ than in the sheet $\hat{\mathbf e}_s$ and sheet--normal $\hat{\mathbf e}_n$ directions and by a nonlinear strain--stress behaviour according to a Fung--type constitutive relation, where the strain energy density reads, 
\begin{equation}
 W_e = \frac{c}{2}(e^Q-1) ,
\end{equation}
with  $Q=\alpha_f \epsilon_{\mathit{ff}}^2+\alpha_s \epsilon_{\mathit{ss}}^2 +\alpha_n \epsilon_{\mathit{nn}}^2 $ being a combination of the Green strain tensor components in the fiber, $\epsilon_{\mathit{ff}}$, sheet, $\epsilon_{\mathit{ss}}$, and sheet--normal $\epsilon_{\mathit{nn}}$ directions.
In the 2D case, the bending stiffness of the tissues is accounted by introducing an additional bending energy potential $W_b =k_b [ 1- \cos(\theta - \theta_0)]$ among two mesh faces sharing an edge, with  $\theta_0$ ($\theta$) their relative free-stress (instantaneous) inclination \cite{kantor1987} and the elastic model parameters have been set as in \cite{fsei}.
\\ \indent 
The left heart is immersed in a Cartesian Eulerian mesh of size $l_x \times l_y \times l_z=120 \times 120 \times 60$~mm$^3$ using the AFiD solver which is based on central second--order finite--differences discretized on a staggered mesh \cite{Verzicco1996,pencil}. The no--slip condition on the wet heart tissues is imposed using an IB technique based on the moving least square (MLS) approach \cite{Vanella2009,TuPa}.
 The hemodynamics is governed by the incompressible Navier--Stokes and continuity equations which in non--dimensional form read:
\begin{equation}\label{NS}
\begin{split}
 \frac{\partial {\bf u}}{\partial t} +  \nabla \cdot  ( {\bf u {\bf u}}) &=  - \nabla p + \nabla \cdot {\bm \tau}  +{\bf f},  \\
\nabla \cdot {\bf u} &= 0,
\end{split}
\end{equation}
where $\mathbf u$ and $p$ are the blood velocity and pressure, whereas $\bm \tau$ if the viscous stress tensor, that in the case of hematic flows in the heart chambers and/or main vessels can be modelled as a Newtonian fluid with the linear constitutive relation 
$\bm \tau =  Re^{-1} (\nabla + \nabla^T ) \mathbf u $. 
An Eulerian grid of $257\times257\times343$ nodes evenly distributed in all three directions is used for the simulations, which has been seen to correctly reproduce the intraventricular hemodynamics and the mitral valve dynamics \cite{fsei,fseichap}.  Since half of a million of time steps are needed to integrate a single heartbeat at 60~bpm, 3 millions timesteps have been integrated to advance 6 heart beats so that to phase--average the results among 5 heart beats, thus discarding the first one. 
In order to provide the hydrodynamic loads as input to the structural solver for fluid--structure coupling, the pressure and the viscous stresses are evaluated at the Lagrangian markers laying on the immersed body surface as
$\mathbf F_f^{ext} =  [ -p_f  \mathbf n_f  +   \bm \tau \cdot \mathbf  n_f ] A_f$, being $A_f$ the area of the triangular face and $\mathbf n$ its normal direction. 
\\ \indent 
The electrical activation of the myocardium is governed by the bidomain model, called in this way because the conductive media is modelled as an intracellular and an extracellular overlapping continuum domains that are separated by the myocytes membrane \cite{tung1978,sundnes2007}. 
The potential difference across the membrane of the myocytes, $v$ the transmembrane potential, and the extracellular potential, $v_{ext}$ extracellular potential, satisfy:
\begin{equation}\label{eq:electro}
\begin{split}
  \chi \left(   C_m \frac{\partial v}{\partial t}  +I_{ion}(\eta) + I_s   \right)  &= \nabla \cdot (\mathcal{M}^{int} \nabla v) + \nabla \cdot (\mathcal{M}^{int} \nabla v_{ext}) ,  \\
    0  &=             \nabla \cdot  (\mathcal{M}^{int} \nabla v + (\mathcal{M}^{int}+\mathcal{M}^{ext}) \nabla v_{ext})),            \\
    \frac{\partial \eta}{\partial t}& = F(\eta,v,t)
\end{split}
\end{equation}
where $\chi$ and $C_m$ are the surface--to--volume ratio of cells and the membrane capacitance. The external triggering stimulus $I_s$ initiates the myocardial depolarization and, since the sinoatrial node that is placed in the upper part of the right atrium is not included in the computational domain, localized $I_s$ are prescribed with the appropriate delays at the Bachmann and His bundles for the left atrium and ventricle, respectively. 
The quantity $I_{ion}$ is the ionic current per  unit cell membrane that is prescribed by the ten~Tusscher--Panfilov model \cite{ten2006} cellular model, indicated by the last equation. 
The parameters $\mathcal{M}^{int} $ and $\mathcal{M}^{ext}$ are the conductivity tensors of the intracellular and extracellular media, which reflect the orthotropic myocardium electrical properties 
with the electrical signal propagating faster along the muscle fiber than in the cross--fibers 
directions \cite{sundnes2007,fsei}. 
The set of equations~\ref{eq:electro} are discretized on the same tetrahedral mesh used for the three--dimensional structural solver by using an in--house finite volume (FV) library, which provides a suitable approach for solving the electrophysiology equation in complex geometries, see \cite{fseichap} where also the electrophysiology parameters are reported. The active muscular tension $ \mathbf F_n^{act}$ at the mesh cell is then obtained as a function of the transmembrane potential $v$ through the model equation proposed by Nash and Panfilov~\cite{nash2004}.
\\ \indent 
The contraction and relaxation of the heart chambers along with the aorta and valve leaflets kinematics results from the dynamic balance between the inertia of the tissues, the external hydrodynamic forces given by the fluid solver $\mathbf F_n^{ext}$, the internal passive forces from the structural solver $\mathbf F_n^{int}$  and the active tension computed by the electrophysiology solver $\mathbf F_n^{act}$:
\begin{equation}\label{eq:newton}
m_n\frac{ \mathrm{d}^2 \mathbf x_n}{\mathrm{d} t^2} = \mathbf F_n^{ext} + \mathbf F_n^{int} +  \mathbf F_n^{act}, 
\end{equation}
where  $\mathbf x_n$ is the (instantaneous) node position and $m_n$ its mass. The hydrodynamics force is non--zero only on the mesh nodes belonging to the wet surfaces (namely the valve leaflets and the inner wall of the chambers), whereas the active tension can be non--zero only for the nodes belonging to the muscular myocardium, i.e. the ventricle and the atrium.

\section{Results}\label{sec:results}
In this section the effects of AS on the left heart flow in terms of hemodynamics, pressure and wall shear stresses is investigated. The healthy configuration with a normal aortic valve of AVA $2.6$~cm$^2$ will be considered in addition to three stenotic cases of increasing disease severity corresponding to (i) mild stenosis with an AVA of $1.8$~cm$^2$, (ii) moderate stenosis with an AVA of $1.1$~cm$^2$ and (iii) severe stenosis with an AVA of $0.5$~cm$^2$.

\subsection{Hemodynamics with a healthy aortic valve}
 \begin{figure}[t!]
\centering
\makebox[\textwidth][c]{\includegraphics[width=1.1\textwidth]{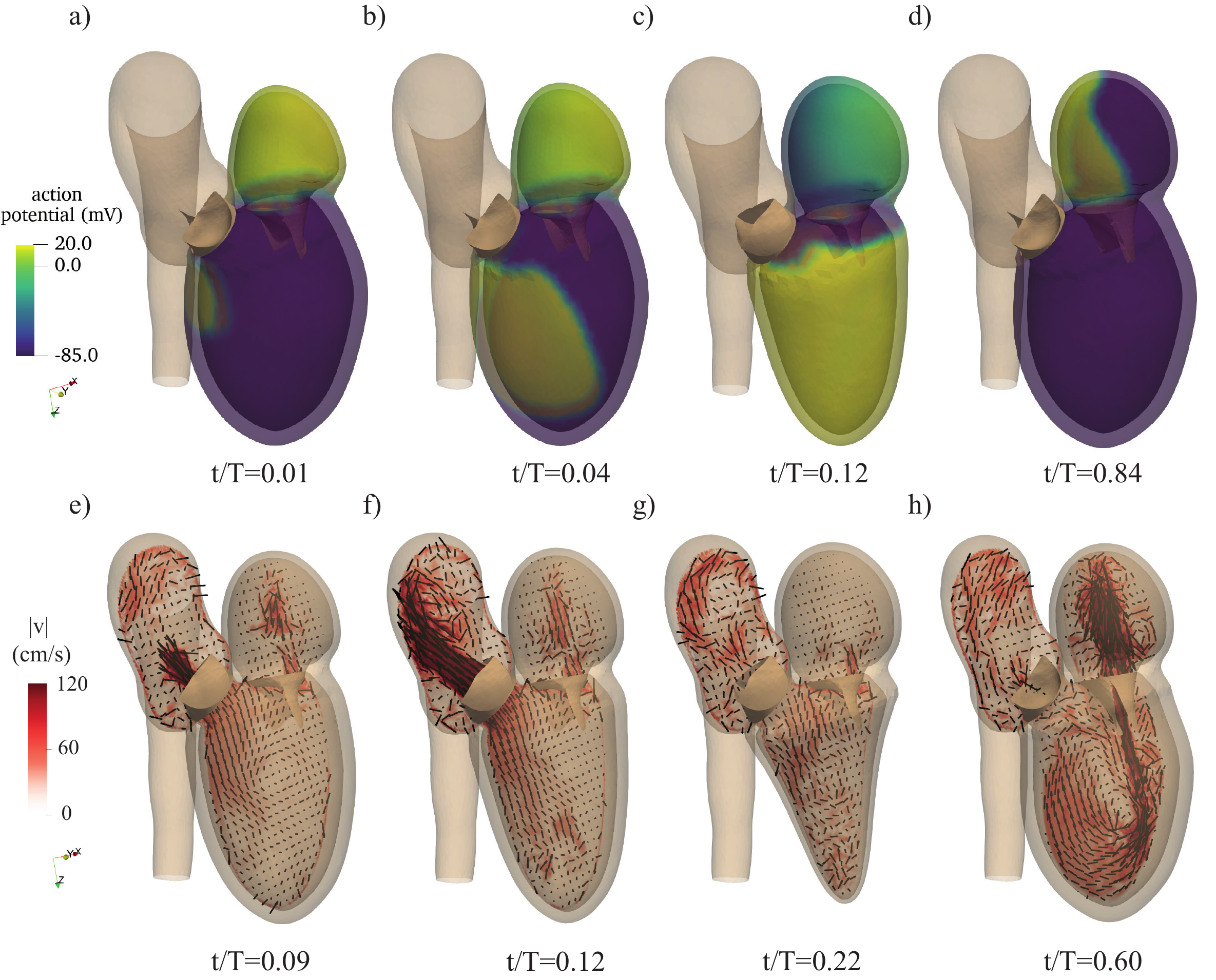}}%
\caption{Fluid--structure--electrophysiology interaction (FSEI) in the healthy case. a-d)  Snapshots of the transmembrane potential propagation in the left heart. e-h) Isocontours of the instantaneous velocity magnitude with superimposed velocity vectors.}
\label{fig:normal}
\end{figure}
Before focusing on the effect of AS on the cardiac hemodynamics, we briefly discuss the case of normal aortic valve that will serve as a reference to better assess the alterations produced by the stenotic pathology.
\\ \indent 
Figure~\ref{fig:normal}(a-d) shows the electrical activation of the left heart. Initially, the myocytes are polarized and the transmembrane potential has the negative value of about $-90$~mV corresponding to the blue isocontour. 
As the electrical impulse is applied at the His bundle (panel \ref{fig:normal}a) the neighboring ventricular myocytes depolarize and reach the positive transmembrane potential of $20$~mV.  This local depolarization results in a propagating wavefront quickly travelling across the tissue (panel \ref{fig:normal}b) and after about 100~ms the whole ventricle is electrically activated (panel \ref{fig:normal}c).
A second electrical impulse is originated at the Bachman bundle, see figure~\ref{fig:normal}(d), and the electrical signal propagates along the atrial muscle causing the cells to locally change the transmembrane potential. 
\\ \indent 
The action potential propagating through the atrial and the ventricular myocardium causes the contraction of the muscular fibers which, while shortening, induce a complex hemodynamics within the heart chambers as visible in Figure~\ref{fig:normal}(e-h), where the hematic flow in the symmetry plane ($x-z$) for representative instants of the heart beat is shown. 
The blood flows  from the upper chamber (the atrium) down to the ventricle that is the main active pump of the heart further propelling the blood in the aorta towards the systemic circulation. 
Owing to the incipient ventricular contraction,  pressure rises at early systole and, when it becomes larger than that of the aorta, the aortic valve opens and blood flows to the aorta, see figure~\ref{fig:normal}(e). As the contraction strengthens the ejected flow rate further increases, thus fully opening the aortic valve leaflets and the maximum blood velocity in the aorta is observed as visibile in panel \ref{fig:normal}(f).
When the systole ends (figure~\ref{fig:normal}g), the ventricle volume is minimum (tele-systolic volume) and its myocytes start to repolarize while the active tension decays yielding the ventricular pressure to decrease and, in turn, the aortic valve closes. 
During early diastole figure~\ref{fig:normal}(h), the ventricle relaxes and the flow accelerates through the mitral orifice thus opening the valve and producing a strong mitral jet (E--wave), which then slows down during diastasis
before another fluid injection generated by the atrial systole creates a second weaker mitral jet (A--wave). These  last two phases of the diastole are not reported here as our investigations will mainly focus on the systole, see \cite{fsei,fseichap} for more results on the diastole for a healthy patient. 

\subsection{Hemodynamics with a stenotic aortic valve}
 \begin{figure}[t!]
\centering
\makebox[\textwidth][c]{\includegraphics[width=1.1\textwidth]{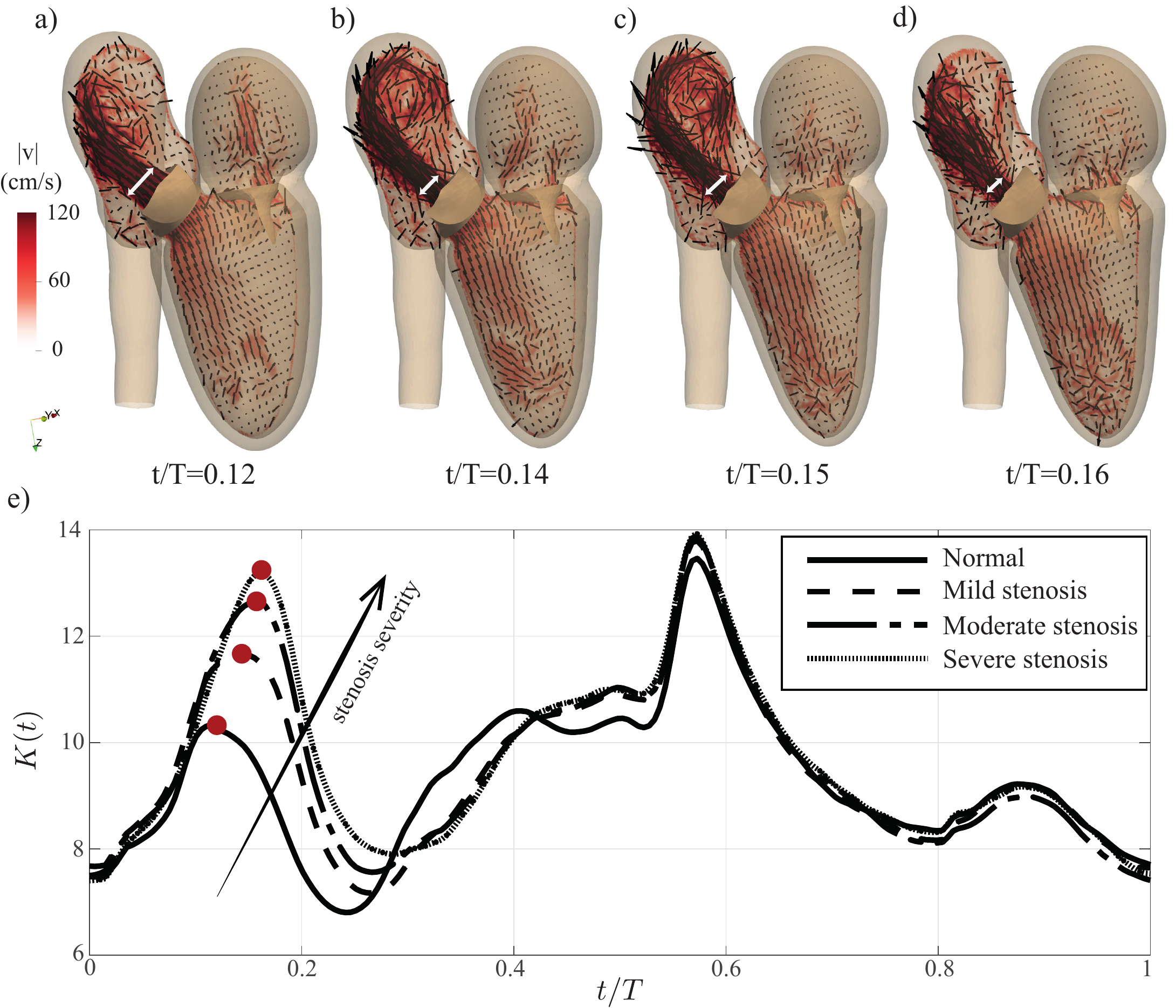}}%
\caption{Isocontours of the instantaneous velocity magnitude with superimposed velocity vectors at peak systole for a) normal valve and b) mild c) moderate d) severe AS. e) Phase--averaged kinetic energy of the flow integrated over the left heart domain as a function of time normalized by the heart beating period.}
\label{fig:jet}
\end{figure}
Although the basic flow features are maintained when the aortic valve leaflets get calcified, some significant alterations emerge during systole, as visible in figure~\ref{fig:jet}(a-d) where the aortic jet at the peak systole for the three stenotic cases is compared against the reference configuration corresponding to the normal aortic valve.
It results that as the AS severity increases, the systolic jet becomes more intense and its section decreases owing to the reduced mobility of the valve leaflets and stronger upward velocities in the aorta take place as indicated by the instantaneous velocity vector (black arrows). Furthermore, the white arrow in the same panels reveals that the diameter of the aortic jet reduces significantly as the AS progresses from mild to severe, accordingly with the reduction of the AVA.
These features are more evident in the sever stenotic case (panel d) where the aortic jet is the narrowest and violently impinges over the inner wall of the aorta.
As the severity of the AS increases the aortic jet becomes also more turbulent with large fluctuations not only  in space but also in time.
\\ \indent 
These hemodynamics alterations due to the stenotic disease also alter the integral kinetic energy per unit mass of the flow 
\begin{equation}
\text{K}(t) = \frac{1}{2}\int_{V{(t)}}  {\bf u} \cdot  {\bf u}~ \mathrm{d} V,
\end{equation}
where the volume $V(t)$ comprises the left ventricle, atrium and aorta.
Figure~\ref{fig:jet}(e) shows that in all cases three energy peaks appear, the first one ($0<t/T<0.3$)  corresponds to the systole and it is generated by the intense velocities at the aortic outflow.
 The second one ($0.35<t/T<0.7$)  is due to the diastolic filling of the ventricle owing to the relaxation of the myocardium (E--wave), whereas  the latter smaller one ($0.85<t/T<1$) is generated by the atrial systole (A--wave).
 As the AS of the aortic valve gets more severe higher values of kinetic energy within the systolic peak are observed, 
whereas the energy peaks occurring during the diastole are basically unaltered as they are due to the mitral valve dynamics which is always the same. 
In particular, the maximum KE with respect to the base level (equal to about 7.4 in nondimensional unit) that is indicated by the red dots increases by 45\% for mild stenosis and by 85\% for severe stenosis with respect to the case of normal aortic valve.
 \begin{figure}[t!]
\centering
\includegraphics[width=.99\textwidth]{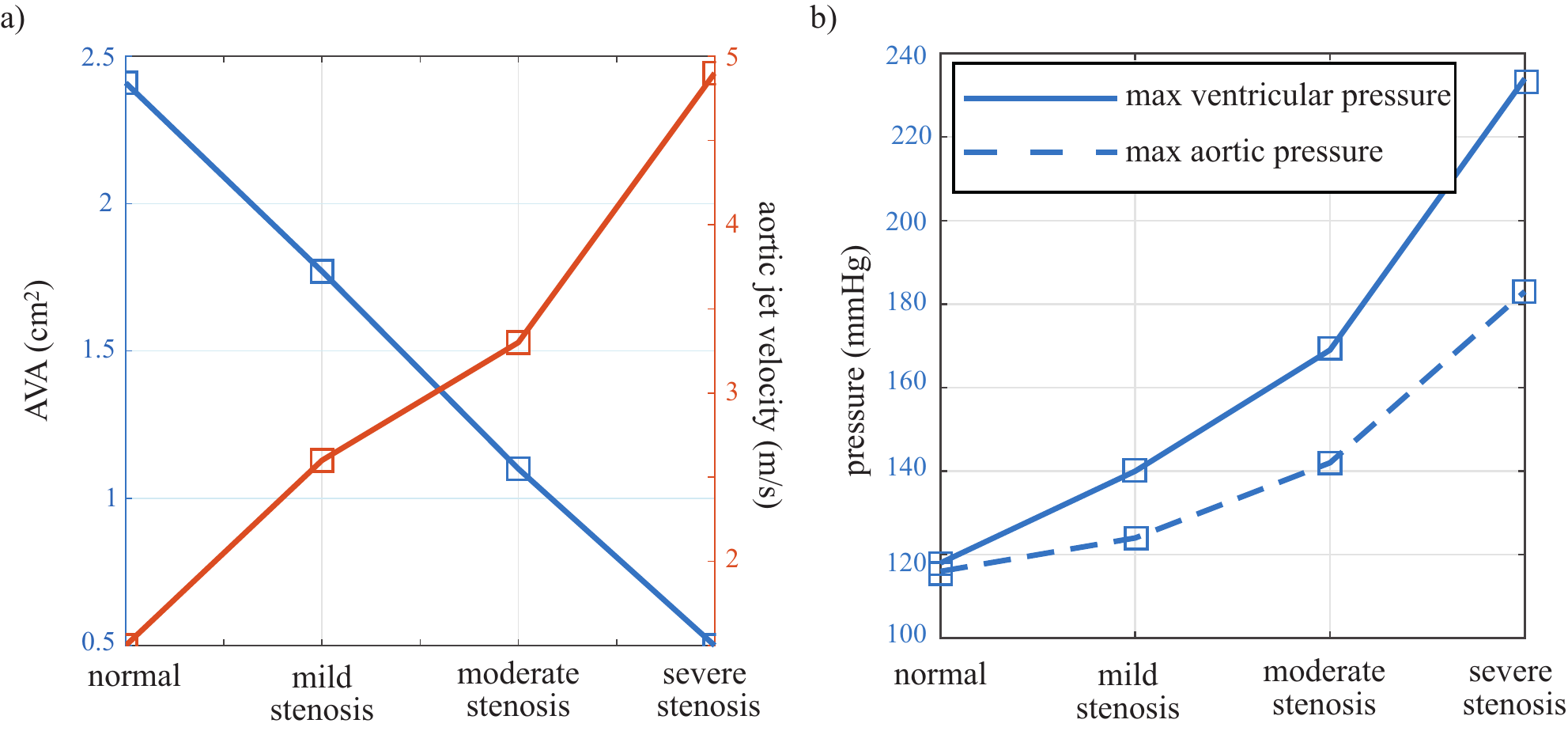}
\caption{a) Aortic valve area (AVA) and aortic jet velocity at peak systole as a function of the AS severity. b) Maximum ventricular and aortic pressure over systole as a function of the AS severity. }
\label{fig:TPD}
\end{figure}
This result is explained by recalling that the aortic jet accelerates as the aortic orifice is occluded by the leaflets calcification and its velocity at peak systole increases from 1.5~m/s in the healthy case to 2.6~m/s for mild stenosis, to 3.3~m/s for moderate stenosis until attaining 4.9~m/s in the case of severe stenosis (i.e. more than three times the normal case), see figure~\ref{fig:TPD}(a).
This speed-up of the aortic jet, however,  is not beneficial for the cardiac flow because it is associated with stronger turbulent fluctuations that may activate hemolysis and produce larger transvalvular pressure drops \cite{rem1,rem2}.   
 \begin{figure}[t!]
\centering
\makebox[\textwidth][c]{\includegraphics[width=1.1\textwidth]{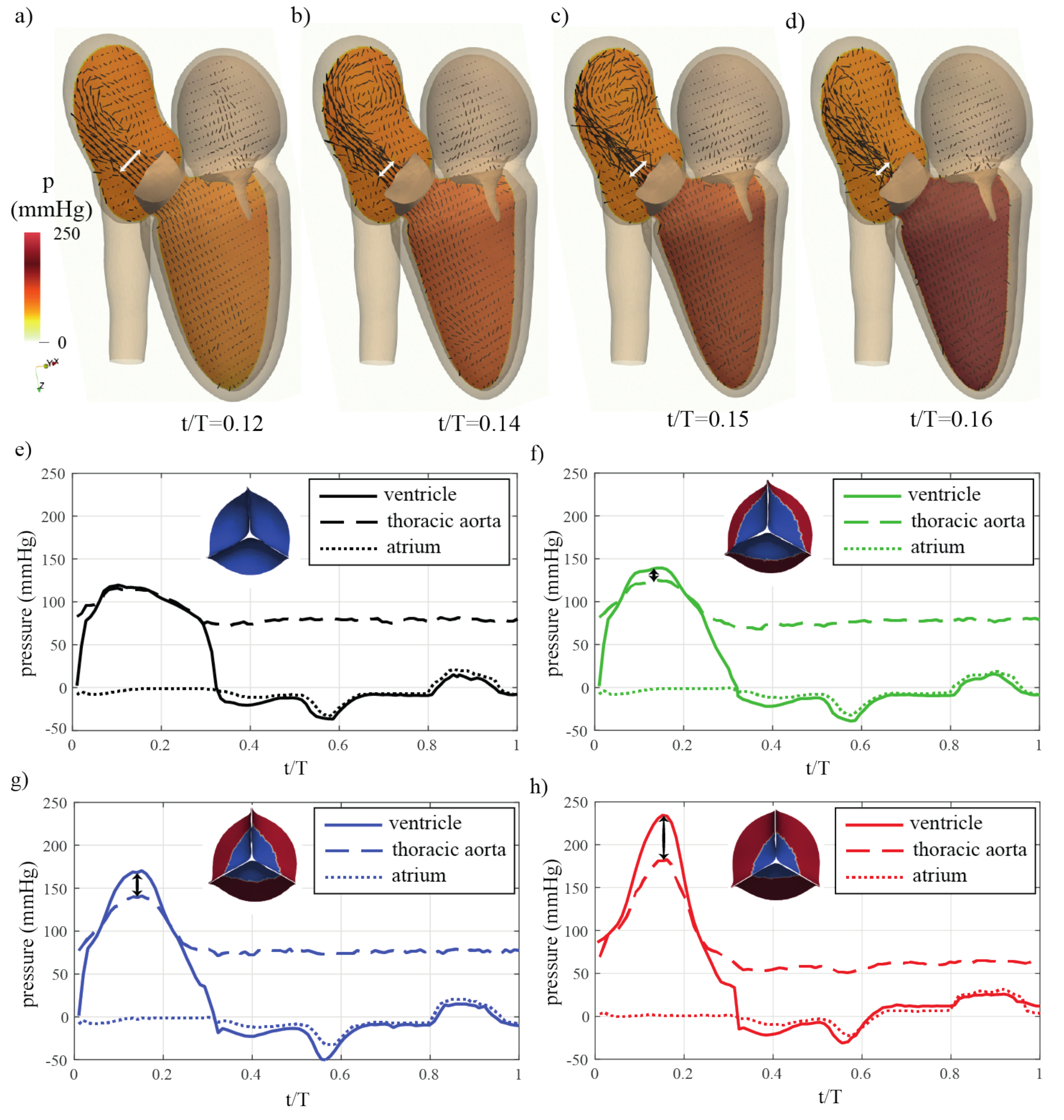}}%
\caption{Isocontours of the instantaneous pressure with superimposed velocity vectors at peak systole for a) normal valve and b) mild c) moderate d) severe AS. Wiggers' diagram for for e) normal valve and f) mild g) moderate h) severe AS.
}
\label{fig:wiggers}
\end{figure}
\\ \indent 
The velocity vectors at the peak systole are also reported in figure~\ref{fig:wiggers}(a-d) now superimposed to the instantaneous pressure field for (a) normal aortic valve along with (b) mild (c) moderate and (d) severe stenosis configurations.
As the stenosis gets more severe, the ventricular pressure increases from 118~mmHg observed in the healthy case to 140~mmHg for mild stenosis to 169~mmHg for moderate stenosis until reaching 234~mmHg for severe stenosis, as also reported in figure~\ref{fig:TPD}(b). 
According to the Laplace law, as the intra--ventricular pressure increases the myocardium wall tension increases proportionally, which may induce long term consequences including heart remodelling \cite{wang2017modelling} as discussed in section~\ref{sec:discussion}. 
 \begin{table}[]
 \centering
 \small
\begin{tabular}{l|cccccc}
         & \multicolumn{2}{c}{ \shortstack{aortic jet  \\ velocity (m/s)}}           & \multicolumn{2}{c}{\shortstack{mean transvalvular \\  pressure (mmHg)}}                                    & \multicolumn{2}{c}{\shortstack{aortic valve \\ area (cm$^2$)}}                                    \\ \hline
         & in-vivo & \multicolumn{1}{l}{numerics} & \multicolumn{1}{l}{in-vivo} & \multicolumn{1}{l}{numerics} & \multicolumn{1}{l}{in-vivo} & \multicolumn{1}{l}{numerics} \\
Normal   &  $<$2.0      & 1.5                & $<5$                           & 2.2                            & $>2.0$             & 2.6                            \\
Mild     & 2.0--2.9      & 2.6                   & 5--20                           & 12.7                            & 1.5--2.0              & 1.8                           \\
Moderate & 3.0--3.9       & 3.3             & 20--40                           & 26.4                          & 1.0--1.5             & 1.1                            \\
Severe   & $\ge 4.0$       & 4.9             & $>40$                           & 42.5                         & $<1.0$              & 0.5                          
\end{tabular}
\caption{Comparison in the classification of aortic stenosis severity between clinical standards \cite{nishimura20172017,baumgartner2009echocardiographic} and present numerical results }
\label{tabella}
\end{table}
\\ \indent 
Since the thoracic aorta and the ventricle are connected through the aortic orifice during systole, also the pressure in the thoracic aorta is altered by the stenotic disease as visible in figure~\ref{fig:wiggers}(e-h) where the Wiggers' diagram is reported. The latter is a standard representation of the heart functioning typically reported in the medical atlas showing the time evolution of pressure in the left--heart chambers: the first peak ($0<t/T<0.3$) corresponds to the overpressure originated by the ventricular systole, whereas the latter smaller one ($0.85<t/T<1$) to the atrial systole. The aortic pressure (dashed line in all panels) has a baseline of about $80$mmHg when the aortic valve is closed, thus uncoupling the two chambers and increases when the aortic valve opens during systole.
As the stenosis level is progressively increased, also the peak systolic pressure in the aorta increases from 116~mmHg measured for normal valve to 124~mmHg for mild stenosis to 142~mmHg for moderate stenosis and attains 183~mmHg for severe stenosis, see~\ref{fig:TPD}(b). Still, according to the Laplace law, the transmural stresses in the aorta increase linearly with the inner pressure in the artery.
\\ \indent 
It should be remarked that not only the systolic pressures in the ventricle and in the atrium increase, but so does their difference corresponding to the TPD, which is a relevant parameter commonly used in the clinical practice to quantify the AS as previously anticipated. 
A mean TPD during systole below 5 mmHg is generally associated to physiological conditions and to a smooth aortic jet, in agreement with the Wiggers' diagram in the healthy case, see figure~\ref{fig:wiggers}(e),
corresponding to a mean TPD of 2.2~mmHg. The same pressure loss raises to 12.7~mmHg in the case of mild stenosis  and to 26.4~mmHg for  moderate stenosis. In the configuration of severe AS a mean TPD of 42.5~mmHg is measured, which may yield to life threatening conditions in the long term.  All quantities obtained from the multi-physics model are well in--line medical observations and are reported in table~\ref{tabella} as a function of the AS severity together with the medical guidelines for the detection of the pathology \cite{braunwald}.

\subsection{Wall shear stress}
 \begin{figure}[t!]
\centering
\makebox[\textwidth][c]{\includegraphics[width=1.0\textwidth]{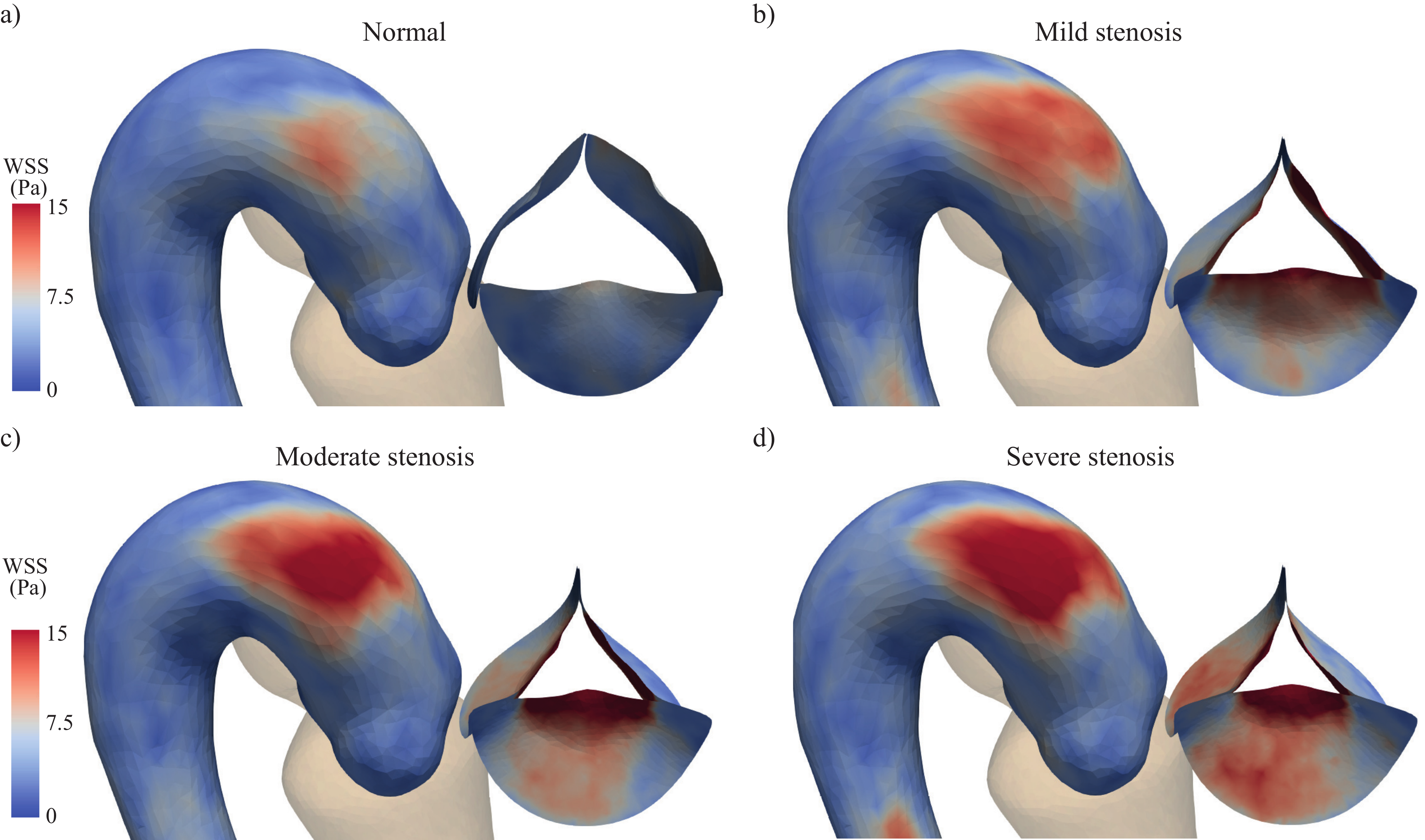}}%
\caption{Distribution of the WSS in the aorta and aortic leaflets for (a) normal valve and (b) mild, (c) moderate, (d) severe AS.}
\label{fig:wss}
\end{figure}
The hemodynamics  alterations originated by the AS have consequences also on the tissue wall shear stress (WSS), which is an important diagnostic parameter of the heart functioning being implied as a cause of disorders through mechanical hemolysis  and tissue remodelling \cite{rem1,rem2}. 
The WSS are determined within the fluid solver and are defined as 
\begin{equation}
\text{WSS} = || \bf t_n - (\bf t_n \cdot \bf n)\bf n ||,
\end{equation}
where $\mathbf n$ the unity vector normal to the surface, $\bf t_n=\bm \tau \cdot \mathbf n$ is the stress vector at the wall with $\bm \tau$ the viscous stress tensor (see section~\ref{sec:fsei}).
Figure~\ref{fig:wss} shows the instantaneous WSS magnitude at the peak systole in the aorta and in the aortic valve leaflets
as a function of the severity of the stenotic disease. 
\\ \indent 
Regarding the aortic leaflets, in the non--stenotic case the WSS is always below 10~Pa with the highest WSS observed at their tips where the shear layers detach merging in the aortic jet  and the orifice area is minimum yielding the blood flow to accelerate due to mass conservation. However, since blood velocity has to match the one of the valve leaflets at their surface (no--slip condition), thin boundary layers are produced over the leaflets corresponding to intense velocity gradients and, in turn, high WSS \cite{fsei}. For mild stenosis (panel b) the systolic jet flowing through the valve is faster than in the healthy case and, consequently, the WSS become more intense not only at the leaflets tip but also over the rest of their surface. The high WSS region further grows for moderate stenosis (panel c) until covering almost completely the leaflet surface for severe stenosis as reported in figure~\ref{fig:wss}(d)
\\ \indent 
As already discussed in figure~\ref{fig:jet}, the systolic jet originated by the ventricular contraction is directed vertically in the aorta and impinges the inner tissue of the aortic arch in the ascending part below its peak.
In the case of healthy aortic valve this location has a small extension, as visible in figure~\ref{fig:wss}(a), and experiences a highest WSS of about 12~Pa. For mild (panel~\ref{fig:wss}b) and moderate (panel~\ref{fig:wss}c) stenosis, the region of high WSS spreads over the aortic arch with maximum values in the order of 14 and 17 Pa, respectively. 
In the case of severe stenosis, the systolic jet further narrows and accelerates thus yielding more intense WSS over a large area of the aortic arch, see panel~\ref{fig:wss}d,  with a maximum WSS of 18 Pa that is 50\%  more intense  than the one measured in the healthy case.
The WSS increase from 10 to 18 Pa might appear inconsequential as the mechanical properties of the aortic tissue can certainly cope with it. However, the endothelium senses the abnormal value of the WSS and, in the medium-- long--term, tissue remodelling is induced as a response to the pathological stimulus which produces irreversible changes in the structure and its functioning.

\section{Discussion}\label{sec:discussion}
In this work the effect of AS on the hemodynamics within the left heart is investigated numerically through a state--of--the--art model which relies on the three way coupling among a Navier-Stokes solver, a structural solver accounting for the orthotropic and hyperelastic tissue mechanics and an electrophysiology solver (bidomain equations). Importantly, the expansion/contraction cycle of the heart chambers and the valve leaflets kinematic are not imposed but come as a part of the numerical solution. 
In particular, the AS induced by calcification is modeled in a similar fashion to \cite{Meschini2019} by preventing the mobility of a certain portion of the aortic leaflets close to the aortic root, where the only control parameter is the radius of the annular region identifying the calcific area setting the severity of the pathology.  In addition to the normal case of healthy aortic valve (all portions of the valve leaflets can move), the model has been set so as to reproduce three different pathological configurations, namely (i) mild, (ii) moderate and (iii) severe stenosis, and for each case the hemodynamics has been solved in order to evidence the abnormal dynamics when the valve pathology occurs.
\\ \indent 
The numerical results evidence that the computational framework and the AS model not only predicts their correct trends as the stenosis severity level increases but also provides data which are consistent with the clinical experience as summarized in table~\ref{tabella}.
In particular, AS is seen to significantly alter the normal hemodynamics in terms of the blood velocity and  pressure, especially during systole when blood is propelled towards the aorta through the aortic orifice by ventricular contraction. 
According to the numerical results, in the normal case with an AVA of 2.6~cm$^2$ the systolic jet has a velocity of $1.5$~m/s at peak systole and a mean TPD during systole of 2.2~mmHg. 
In the stenotic case the systolic jet becomes faster and more irregular owing to the reduced mobility of the valve leaflets with aortic jet velocity at peak systole increasing up to 2.6~m/s for mild stenosis and to 3.3~m/s for moderate stenosis, until getting to 4.9~m/s for severe stenosis that is more than three times the reference case of normal valve. Accordingly, the mean TPD during systole increases from 12.7 to 42.5~mmHg  as the stenosis severity is increased from mild to severe with important implications on the transmural loads withstood by the heart chambers which are known to increase proportionally. 
\\ \indent
A great advantage of  the numerical approach proposed here is the possibility to quantify the alteration on the WSS induced by the AS, which can not be measured directly in--vivo. For both healthy and pathological cases, the highest WSS occur at the initial ascending tract where the aortic jet hits the inner artery wall, but as the stenosis gets more severe the narrower and faster systolic jet yields both an higher WSS level and a larger surface distribution. A similar trend is also observed on the surface of the valve leaflets where 
the WSS are normally more intense at their tips. As AS progresses the systolic flow accelerates, owing to a reduction of AVA and the region of high WSS gets not only more intense, but also spreads all over the surface of aortic leaflets including the annular outer region already calcified.
This altered WSS distribution may have  long term consequences on the remodelling of the structures and may start irreversible diseases changing the structures of the heart according to a dilatative cardiomyopathy, or a hypertrophic remodeling \cite{braunwald,wang2017modelling}.
\\ \indent 
Furthermore, in the case of severe stenosis the augmented shear stresses and turbulent intensity within the systolic phase  
can also damage red blood cells thus inducing platelet activations, which eventually leads to thromboembolic events arising from the formation of clots and their subsequent detachment \cite{smith1972thrombus,stein1974measured,de2009direct}. 
These complications require lifelong anti--coagulation therapy, inducing high level of hemorrhage risk.
Therefore, a natural continuation of this work would be to measure the damage of the red blood cells as a function of the AS severity by considering not only the Eulerian computation of the Reynolds stresses, but also tracking the trajectories of fluid particles to measure the instantaneous viscous stress tensor and the exposure time along their paths  \cite{de2009direct}. Such a Lagrangian approach would allow to better quantify the red blood cells damage and predict the onset of the hemolysis process as a function of the severity of the stenotic disease. 
A possible improvement of the computational model would be to consider a non--uniform thickness distribution for the valve leaflets with a progressive increase of the tissue thickness approaching the calcific annular region of the valve instead that a uniform thickness as considered in the present work. Furthermore, a simulation campaign could be run by varying the thickness and the stiffness of the valve leaflets, which are known to be correlated with the patient age as the degeneration of collagen fibers and to lipid accumulation and calcification, the leaflet thickness increases with ageing \cite{sahasakul1988age}.

\section*{Acknowledgments}
This work has been partly supported with the
865 Grant 2017A889FP 'Fluid dynamics of hearts at risk of failure: towards methods for the prediction of disease progressions' funded by the Italian Ministry of Education and University.

\section*{References}

\bibliography{mybibfile}

\end{document}